\newcount\Comments
\newcount\ResolvedComments
\documentclass[letterpaper]{article}

\usepackage{aaai}
\usepackage{times}
\usepackage{helvet}
\usepackage{courier}

\usepackage{amsmath,amsfonts}
\usepackage{amssymb}
\usepackage{comment}
\usepackage{textcomp}
\usepackage{verbatim}

\usepackage{graphicx}
\usepackage{tikz}
\usetikzlibrary{patterns,shadows,arrows,decorations,decorations.shapes,backgrounds,shapes}

\usepackage[font=small,format=plain,labelfont=bf,up]{caption}
\DeclareCaptionType{copyrightbox} 
\usepackage{subfig}
\usepackage{multicol}

\usepackage{enumitem} 

\setitemize{noitemsep,topsep=3pt,parsep=0pt,partopsep=0pt}

\newlength{\figurewidth}
\makeatletter
\renewcommand{\paragraph}{%
  \@startsection{paragraph}{4}%
  {\z@}{0.25ex \@plus 1ex \@minus .2ex}{-1em}%
  {\normalfont\normalsize\bfseries}%
}
\makeatother

\newcommand{\citet}[1]{\citeauthor{#1}~(\citeyear{#1})}

\newcommand{\ignore}[1]{}
\newcommand{\kibitz}[2]{\ifnum\Comments=1\textcolor{#1}{#2}\fi}

\newcommand{\am}[1] {\kibitz{red}{\bf\noindent [AM: #1]} }
\newcommand{\yc}[1] {\kibitz{teal}{\bf\noindent [YC: #1]} }

\newcommand{\resolved}[1] {\ifnum\ResolvedComments=1\textcolor{black}{[#1]}\fi}

\newcommand{\E}{\mathbf{E}}
\newcommand{\A}{\mathcal{A}}
\newcommand{\M}{\mathcal{M}}
\newcommand{\N}{\mathcal{N}}
\renewcommand{\P}{\mathcal{P}}
\newcommand{\LO}{\mathcal{L}}

\newcommand{\etal}{{\it et al.} }

\begin{document}

%
\title{Capturing Variation and Uncertainty in Human Judgment}


\author{
  Andrew Mao \\ 
  Harvard University \\ 
  mao@seas.harvard.edu 
  \And
  Hossein Azari Soufiani \\
  Harvard University \\ 
  azari@fas.harvard.edu
  \And
  Yiling Chen \\ 
  Harvard University \\ 
  yiling@eecs.harvard.edu
  \And
  David C. Parkes \\ 
  Harvard University \\ 
  parkes@eecs.harvard.edu
}


\maketitle


\begin{abstract}
  The well-studied problem of statistical rank aggregation has been
  applied to comparing sports teams, information retrieval, and most
  recently to data generated by human judgment. Such human-generated
  rankings may be substantially different from traditional statistical
  ranking data. In this work, we show that a recently proposed
  generalized random utility model reveals distinctive patterns in
  human judgment across three different domains, and provides a
  succinct representation of variance in both {\it population
    preferences} and {\it imperfect perception}.
  In contrast, we also show that classical statistical ranking models
  fail to capture important features from human-generated input. Our
  work motivates the use of more flexible ranking models for
  representing and describing the collective preferences or
  decision-making of human participants.
%
%
\end{abstract}

\section{Introduction}
\label{sec:intro}

The problem of creating a ranking from noisy ranking or pairwise
comparison data is widely studied across statistics, economics and
machine learning. Statistical rank aggregation has been used to rank
sports teams and racing drivers~\cite{Stern1990}, for information
retrieval~\cite{Liu2009}, for collective ideation~\cite{Salganik12},
in preference learning~\cite{Kamishima03}, and for social
choice~\cite{CRX09}.

Common to many current applications is that the input data comes from
people, such as preferences in collaborative filtering or
crowdsourcing and quality judgments in human computation. In
crowdsourcing, a common problem is that of choosing the best or most
desirable alternative from a large set of possibilities using ranking
or voting~\cite{Chen13}. For example, \citet{LCGM10} give an example
of ranking a list of suggestions for things to do in New York.
%

Yet, human-generated ranking data can be fickle, encompassing both
varying preferences between users and imperfect perception or
judgment, and simply producing an aggregate ranking overlooks nuances
in the data. Instead, we propose looking beyond simple rank
aggregation to also understand the patterns of decision making that
emerge. Specifically, we are interested in viewing human-generated
rankings under one or both of the following settings:

\begin{itemize}
\item {\bf Population preference}: Different rankings arise from a
  distribution over the population. By learning this distribution, we
  discover the most common preferences in the population as well as
  how uniform or varied preferences are across users.
\item {\bf Imperfect perception}: The variability in rankings arise
  from errors in the perception of an underlying truth. By learning
  this distribution, we discover which comparisons are noisy or
  cognitively difficult for users.
\end{itemize}

We consider three human-generated data sets demonstrating these two
settings: preferences over types of sushi, decisions about ranking
$8$-puzzles by distance to the solution state, and decisions about
ranking pictures of dots by number. Using several probabilistic models
of rankings, we show that a generalized random utility model (RUM)
based on the normal distribution~\cite{AzariNIPS12}, is able to better
explain the variability in the data than the classical and often used
Mallows~\shortcite{Mallows57} and
Plackett-Luce~\shortcite{Luce59,Plackett75} models.

In particular, we show that the Normal RUM is significantly better in
matching the empirical pairwise comparison probabilities---the
marginal probability that one alternative is ranked ahead of
another---in the data, and reveals interesting patterns as a result.
In data over preferences, we discover users' ubiquitous affinity or
dislike for certain alternatives as well as distinguishing between
conventional and more controversial items. In data about decision
making, we reveal the comparisons are that harder or easier to make,
and how the difficulty of a ranking task affects these comparisons.
In contrast, we also demonstrate why the Mallows and Plackett-Luce
models have inherent limitations in capturing heterogeneity in
human-generated data. We believe that these insights derived from more
flexible models will prompt the use of new techniques for describing
human preferences and perception beyond simple rank aggregation.

In the rest of the paper, Section~\ref{sec:models} provides an
overview of statistical ranking models. Sections~\ref{sec:sushi}
and~\ref{sec:ranking} show how the Normal RUM allows for the
interpretation of collective behavior on data where other ranking
models do not (Appendix~\ref{sec:model-properties} analytically
explains why the classical models fail to reveal this behavior.)
Section~\ref{sec:discussion} discusses implications and future work.

\section{Ranking Models}
\label{sec:models}


We consider ranking problems with orderings over $m$ alternatives
provided by $n$ agents. Let $\A = \{a_1, a_2, \dots, a_m\}$ denote the
set of alternatives. Let $\LO$ denote the set of all total orders over
$\A$. Let $\M$ with parameters $\theta$ denote a ranking model,
generating an i.i.d. distribution on rankings: $\sigma_i \sim
\M(\theta), \sigma_i \in \LO, i \in \{1 \ldots n\}$ denotes the
$i$\textsuperscript{th} ranking in the data. 
%
%
%
%

In a {\it random utility model}~\cite{Thurstone27,McFadden1974},
each alternative $a_j$ has a random value (or utility) $x_{j} = \mu_j
+ \epsilon_{j}$, where $\epsilon_{j}$ is a zero-mean noise component,
usually independent across alternatives, and $\mu_j\in \mathbb{R}$ is
the mean.
%
%
The realized values $(x_1,\ldots,x_m)$ induce a ranking $\sigma$ with
$a_j\succ a_k \ \Leftrightarrow\ x_j>x_k$.
%

Different distributions for $\epsilon_{j}$ correspond to different
random utility models (RUMs).
%
In the Normal RUM~\cite{AzariNIPS12}, 
%
$\epsilon_{j} \sim \N(0, \sigma_j^2)$, where $\sigma_j$ can vary
across alternatives. Although a straightforward model, it has been
historically intractable to use; Azari \etal addressed this by
adopting Monte Carlo Expectation Maximization (MC-EM) to estimate the
parameters.
%
%


The classical {\em Plackett-Luce} model~\cite{Luce59,Plackett75} can
be interpreted as a RUM in which the noise terms $\epsilon_{j}$ are
independent Gumbel distributions with different means and the same
variance~\cite{Yellott77}.
%
The Plackett-Luce model is 
popular due to its tractability. In particular, the likelihood
function has a simple closed form and can be optimized efficiently
with algorithms such as {\em
  minorization-maximization}~\cite{Hunter04}.
%

Another popular ranking model is the {\em Mallows}
model~\cite{Mallows57} in the statistical literature (also called {\em
  Condorcet's model} in social choice.)  Mallows is not a random
utility model. Rather, the parameters $\theta$ define a {\em reference
  ranking} $\sigma^*\in {\mathcal L}$ and a {\em noise parameter} $p
\in (0.5, 1]$. This model generates a random ranking by ordering all
ordered pairs $(a_j, a_k)$ in agreement with reference $\sigma^*$ with
probability $p$, and disagreement otherwise. If the result is an
(acyclic) rank order than it is retained; otherwise, the process is
repeated.  A challenge with the Mallows model is that estimating the
maximum likelihood parameters is NP-hard---in fact, the MLE of the
reference ranking parameter is the social rank determined by the
Kemeny voting rule~\cite{Young88}.

\label{sec:model-extensions}

Many extensions have been proposed to the above models in attempts to
describe heterogeneity in data, such as by modeling agent-specific
features or correlation between alternatives. The Mallows model has
been extended to allow for mixtures~\cite{LB11b}, and
there are many extensions to the Plackett-Luce model, such as
\cite{Xia2008,Qin2010}, which allow for a mixture of distributions or
for the random utility parameters to depend on other features.
The goal of these extensions has been largely to increase the models'
descriptiveness. In this work, we focus on simple models assuming that
people are {\it ex ante} symmetric, and show that the Normal RUM
already reveals interesting new observations about human judgment.

We implement maximum likelihood estimation for all the models
previously described using multi-core parallelization, with numerical
optimizations where possible. 
%
Our code is available as an integrated
package\footnote{https://github.com/mizzao/libmao} with the aim of
increasing the accessibility of using many different algorithms for
modeling rank data, allowing for analysis and visualization using the
methods we show in this work.

\section{Sushi Ranking Dataset}
\label{sec:sushi}

\begin{figure*}[t!]
  \centering
  \subfloat[Mallows pairwise probabilities. Negative log likelihood: 71353.]  {
    \includegraphics[width=0.33\textwidth]{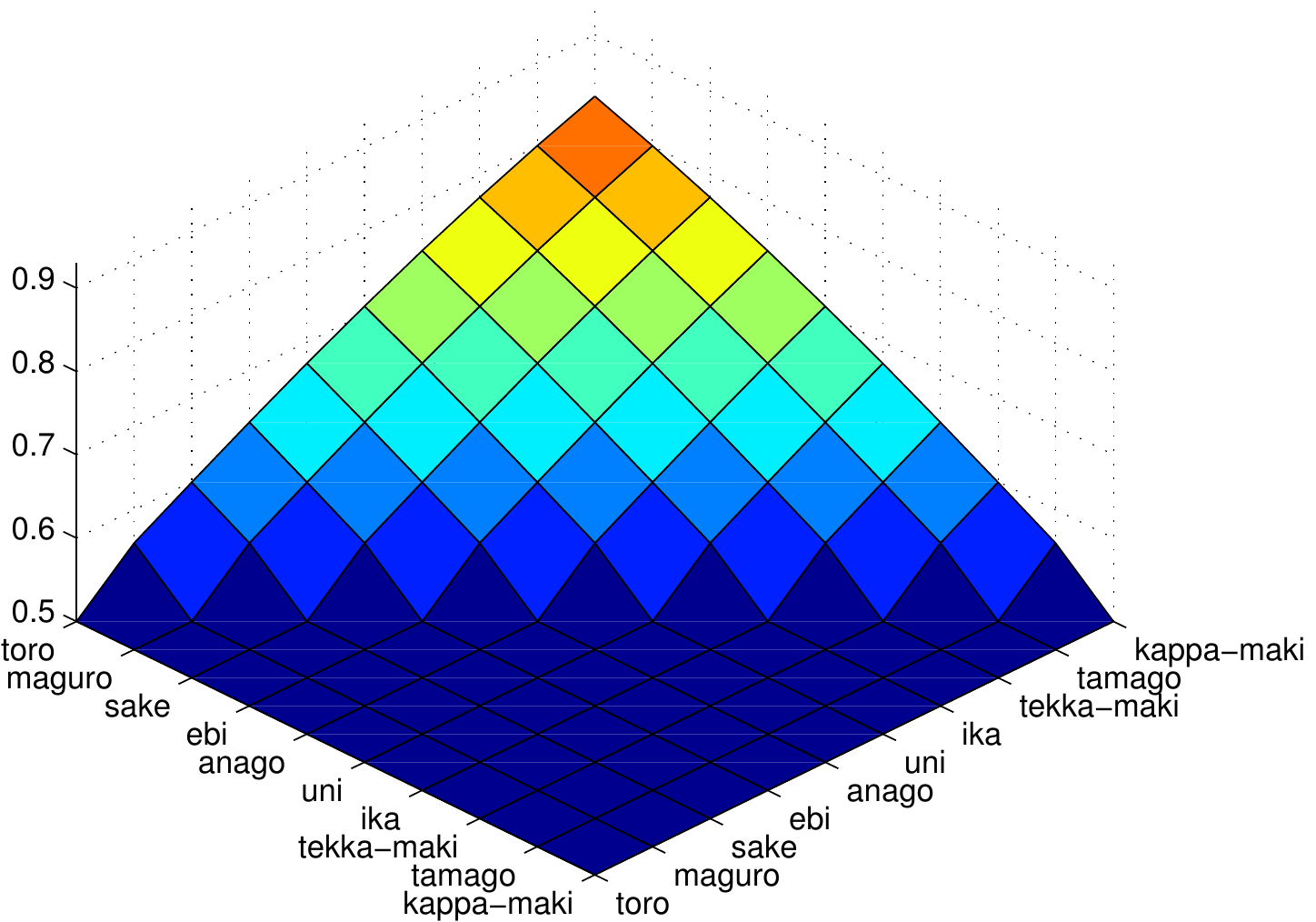}
    \label{fig:mallows-pairwise}
  }
  \subfloat[Empirical probabilities in order.]  {
    \includegraphics[width=0.33\textwidth]{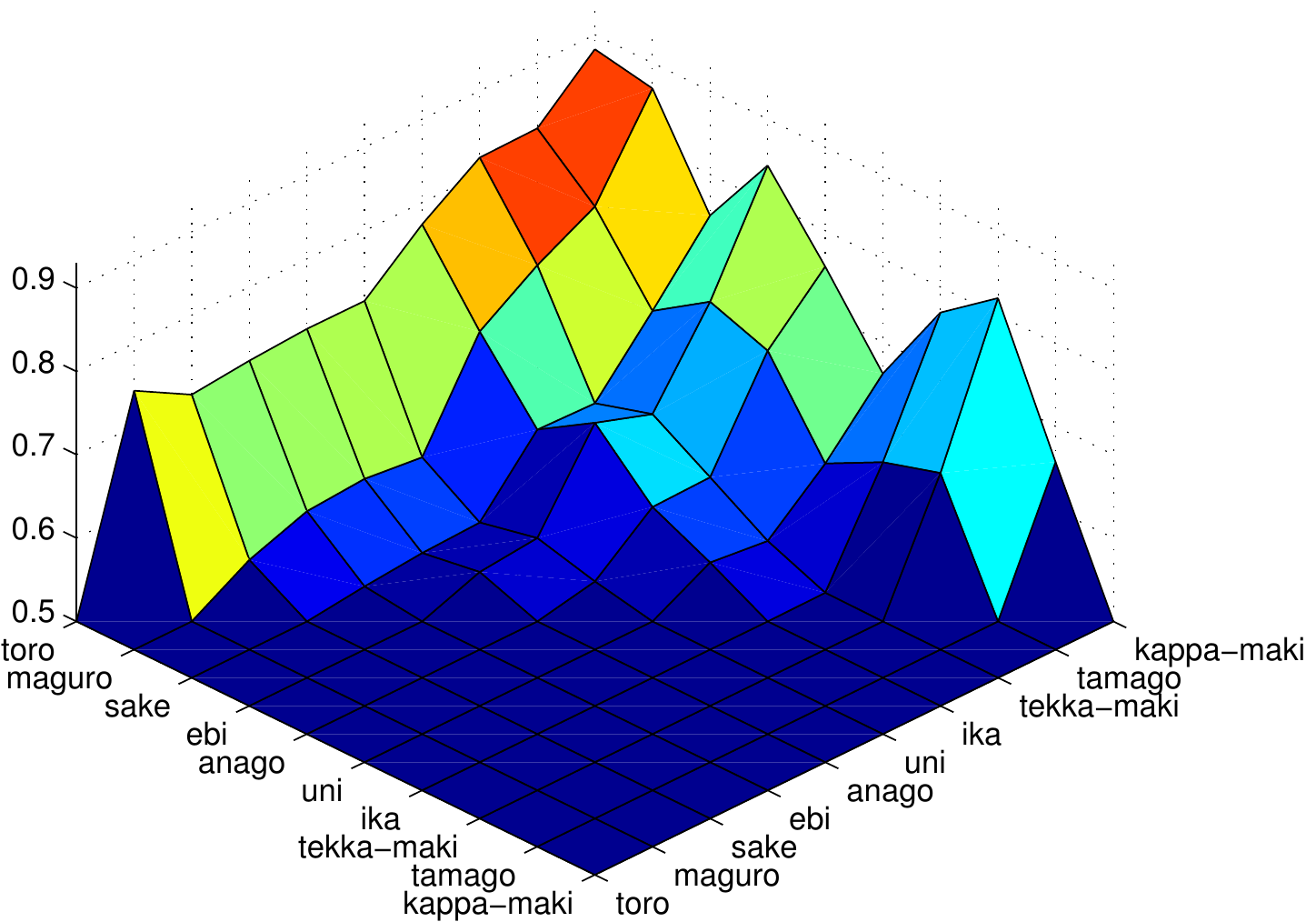}
    \label{fig:mallows-empirical-pairwise}
  }
  \subfloat[Mallows deviation.]  {
    \includegraphics[width=0.33\textwidth]{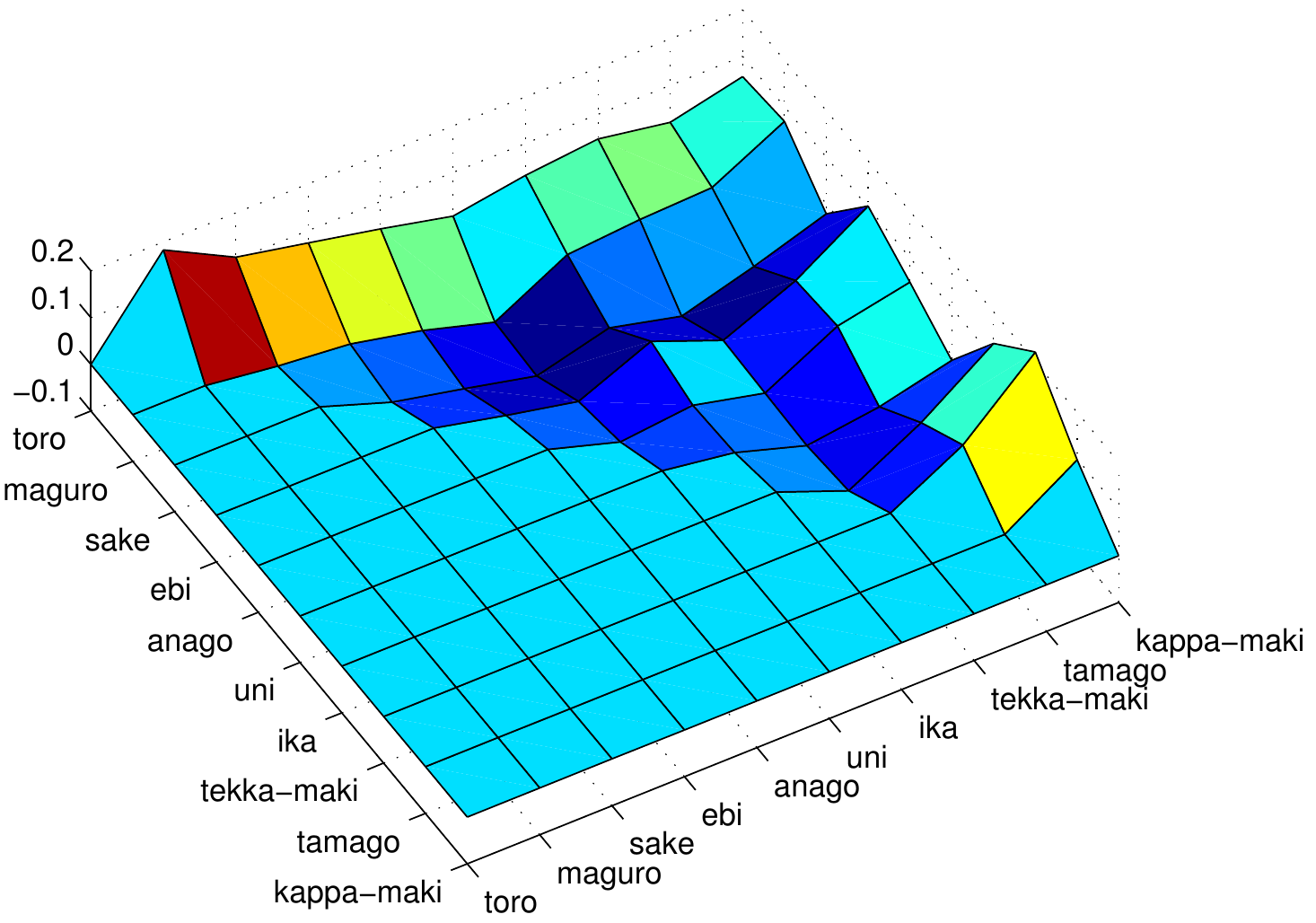}
    \label{fig:mallows-deviation}
  }
  \\ \vspace{-10pt}
  \subfloat[Plackett-Luce pairwise probabilities. Negative log likelihood: 71211.]  {
    \includegraphics[width=0.33\textwidth]{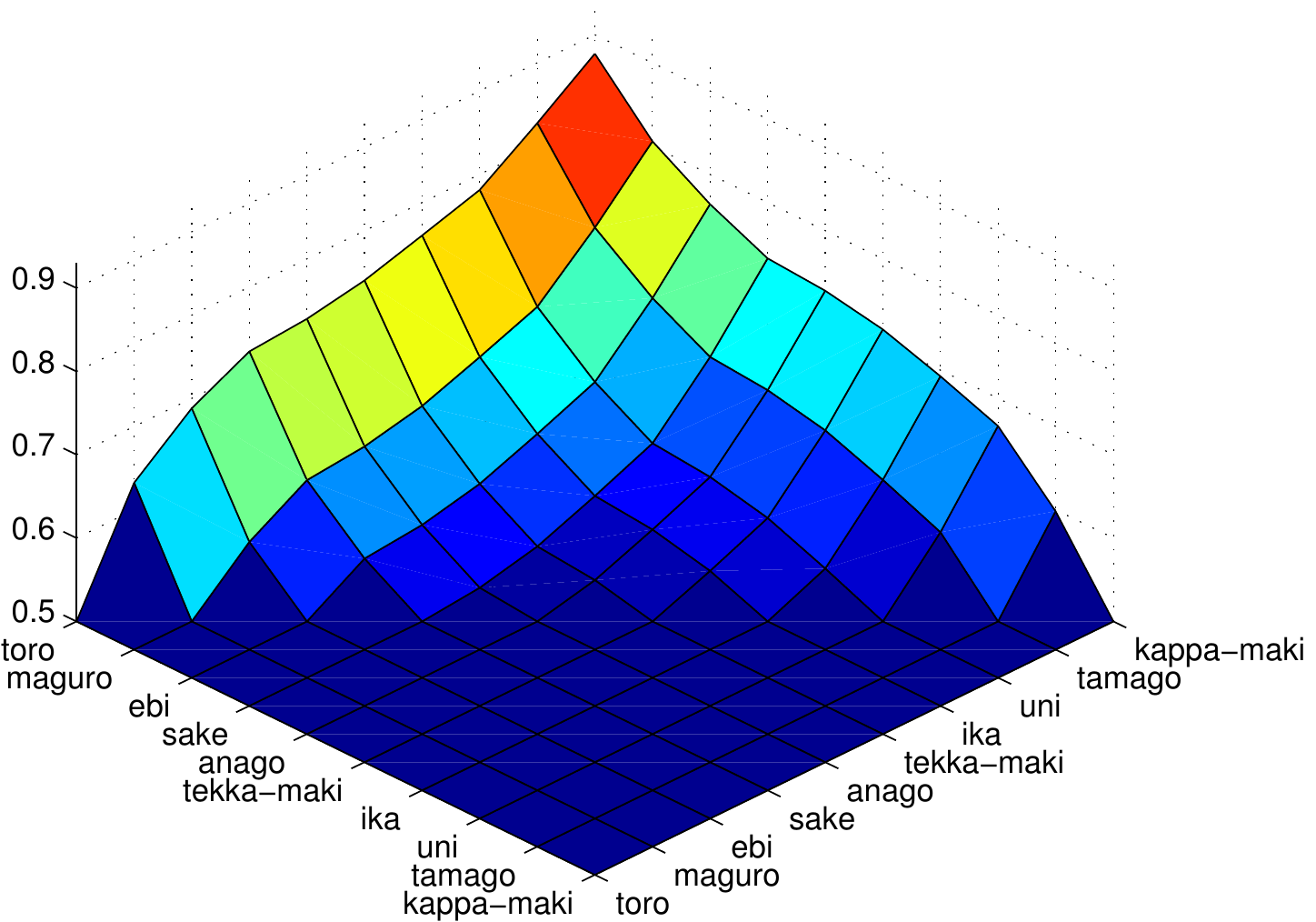}
    \label{fig:pl-pairwise}
  }
  \subfloat[Empirical probabilities in order.]  {
    \includegraphics[width=0.33\textwidth]{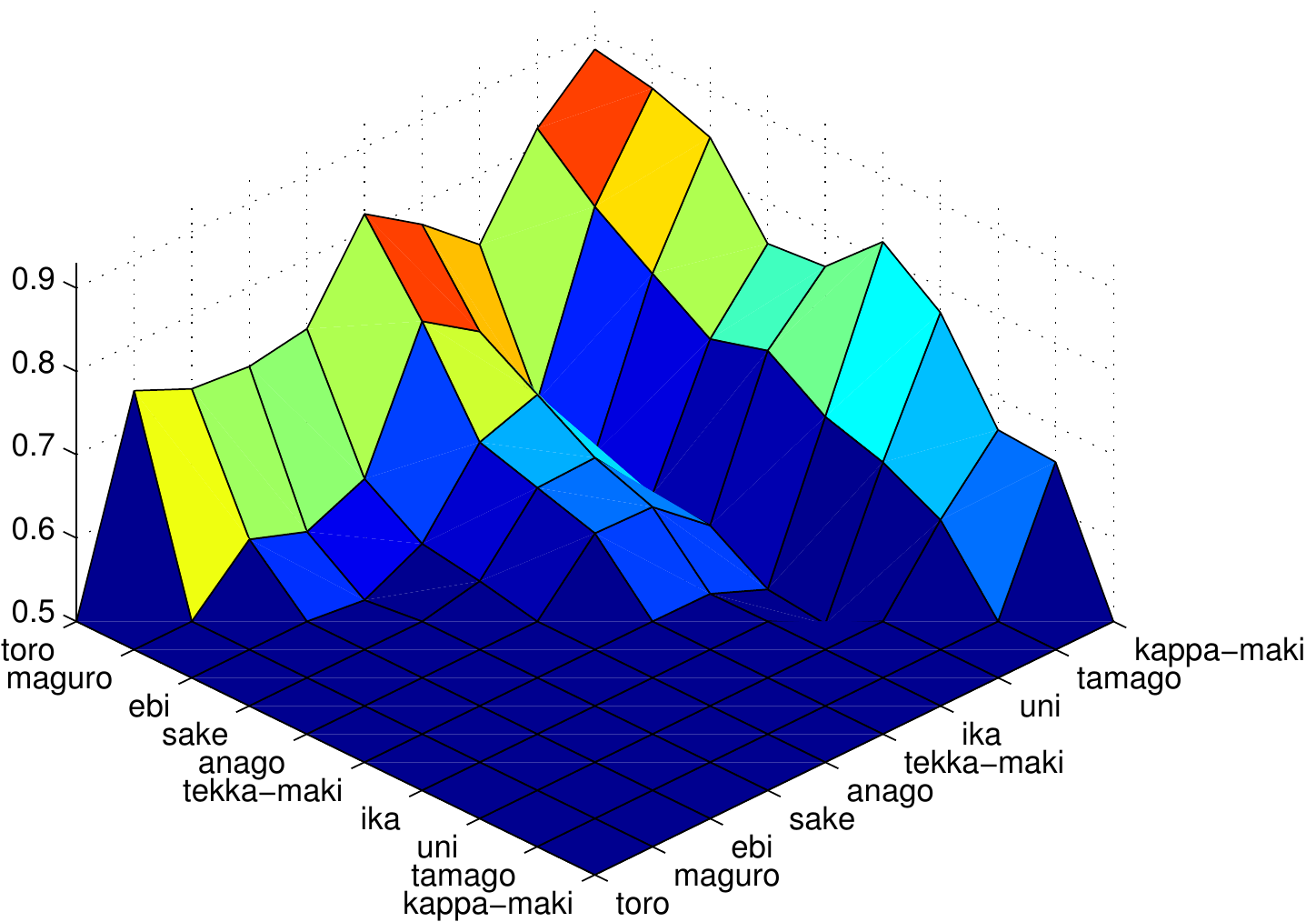}
    \label{fig:pl-empirical-pairwise}
  }
  \subfloat[Plackett-Luce deviation.]  {
    \includegraphics[width=0.33\textwidth]{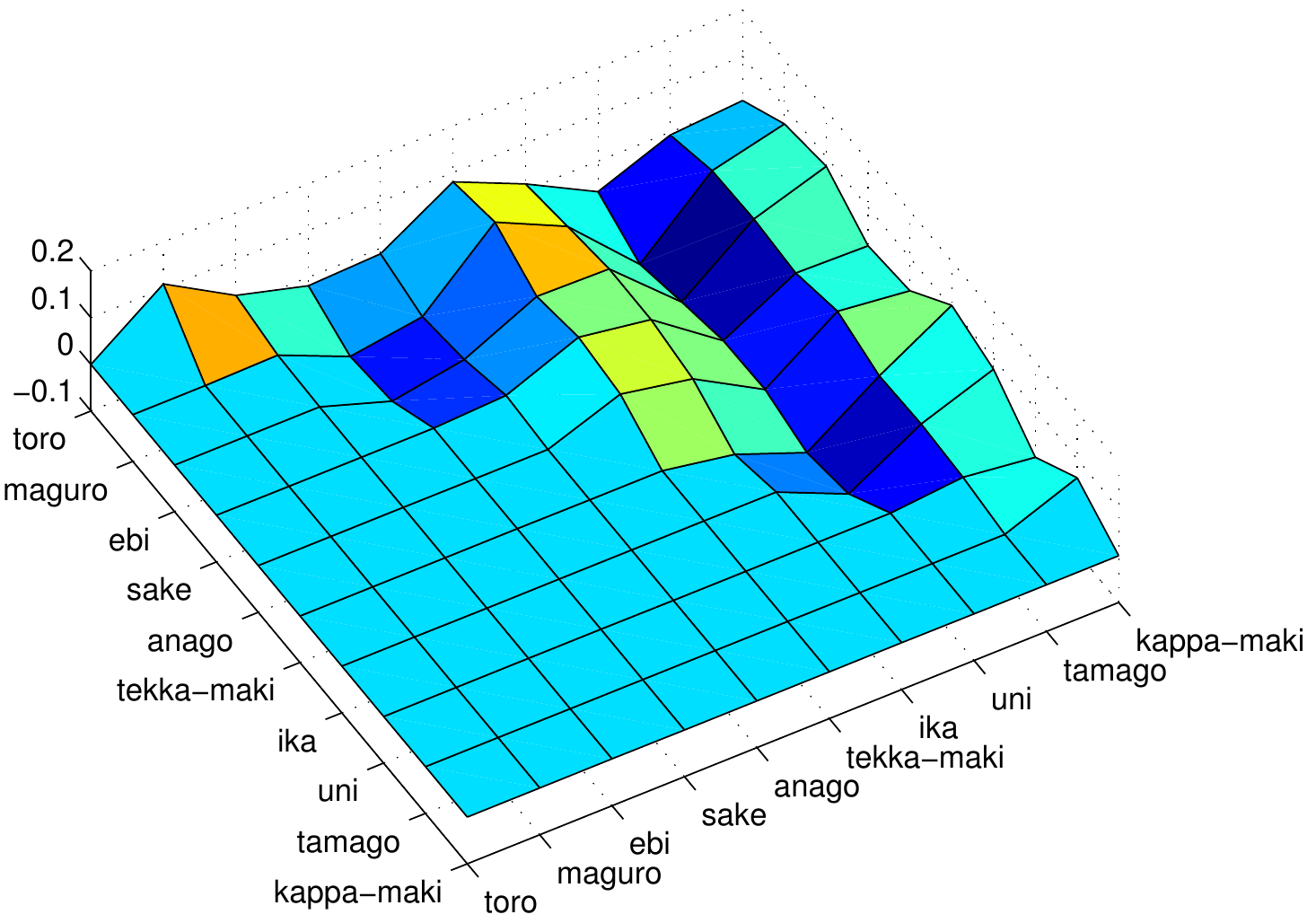}
    \label{fig:pl-deviation}
  }
  \\ \vspace{-10pt}
  \subfloat[Normal RUM pairwise probabilities. Negative log likelihood: 69011.]  {
    \includegraphics[width=0.33\textwidth]{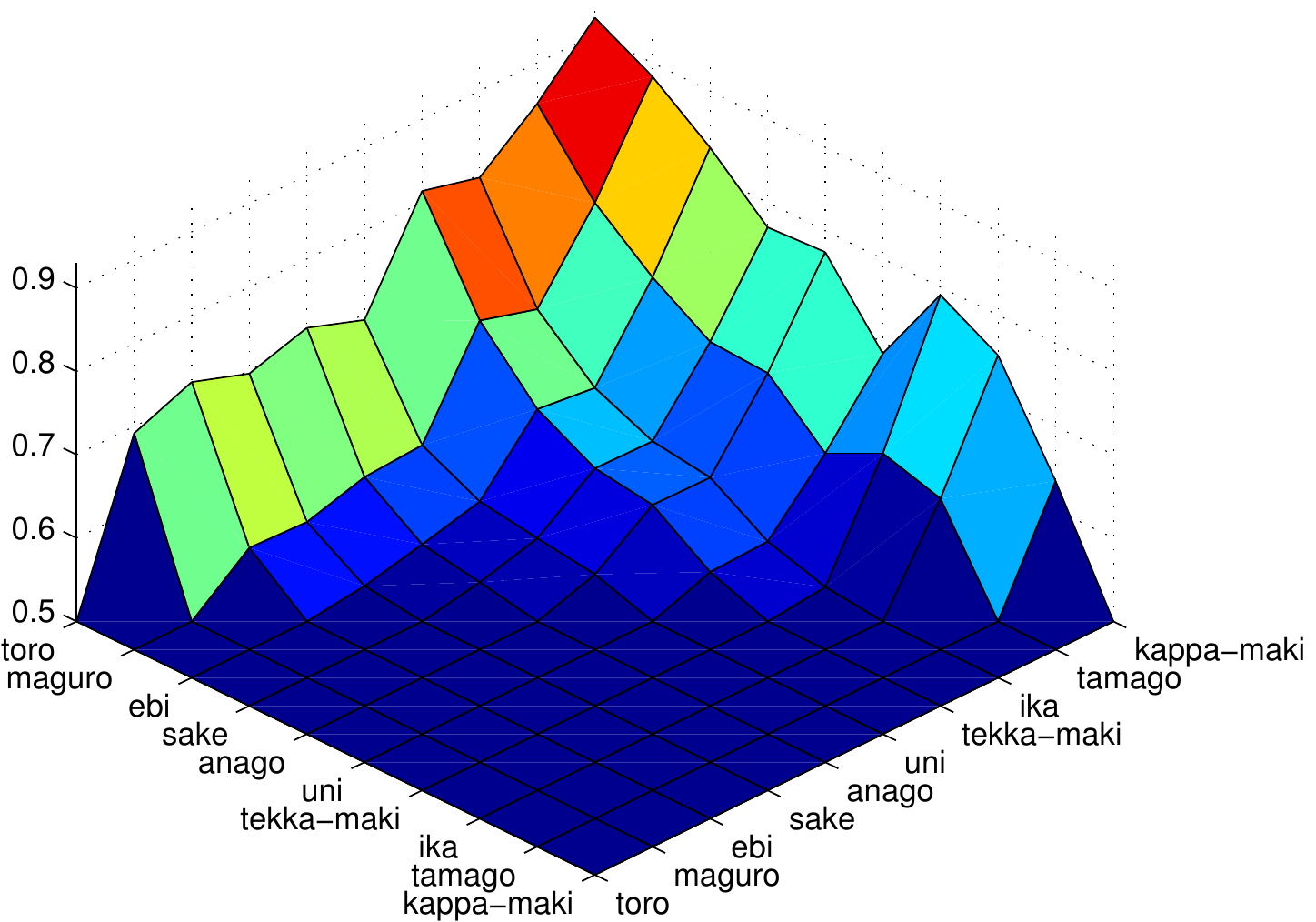}
    \label{fig:normal-pairwise}
  }
  \subfloat[Empirical probabilities in order.]  {
    \includegraphics[width=0.33\textwidth]{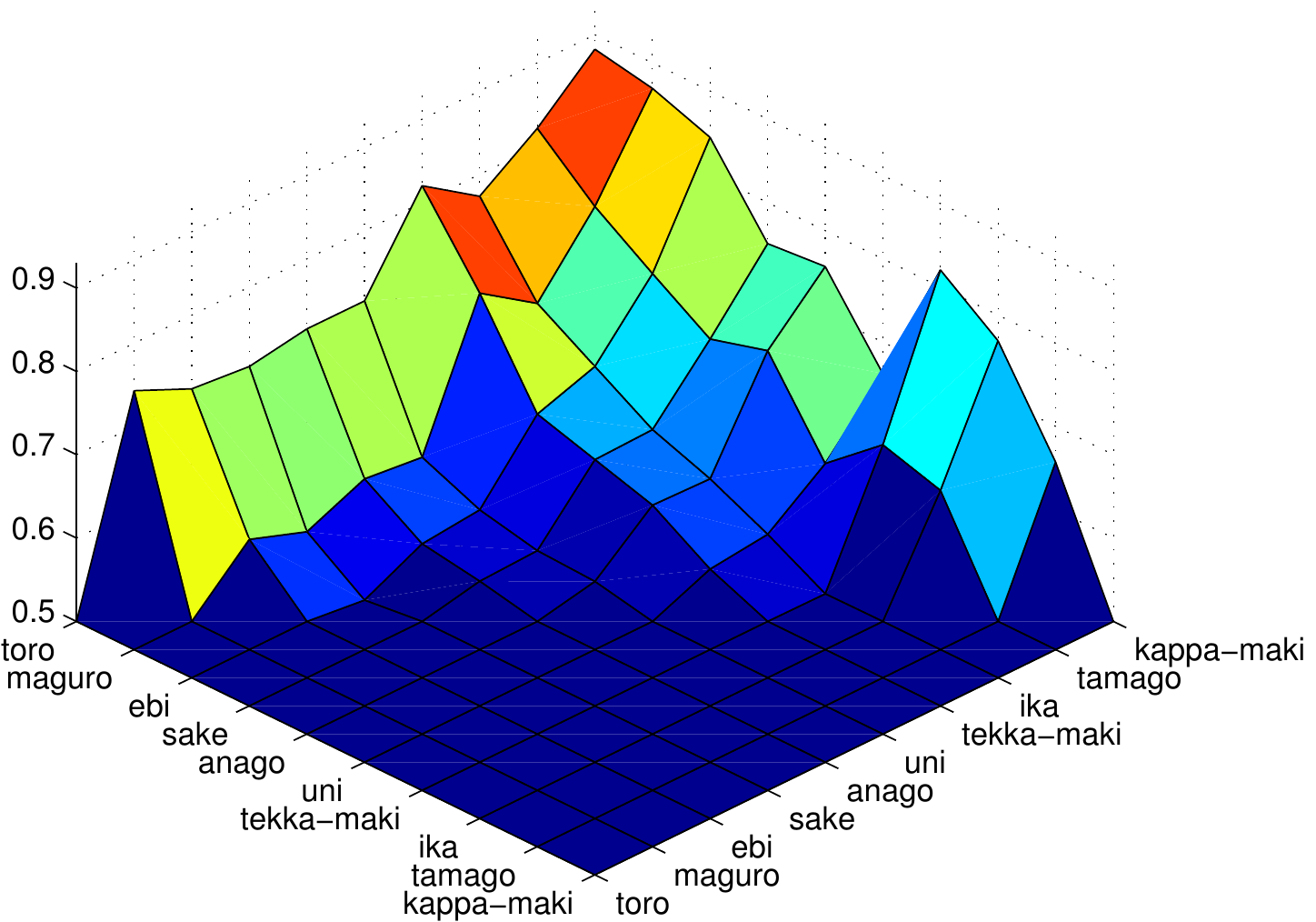}
    \label{fig:normal-empirical-pairwise}
  }
  \subfloat[Normal RUM deviation.]  {
    \includegraphics[width=0.33\textwidth]{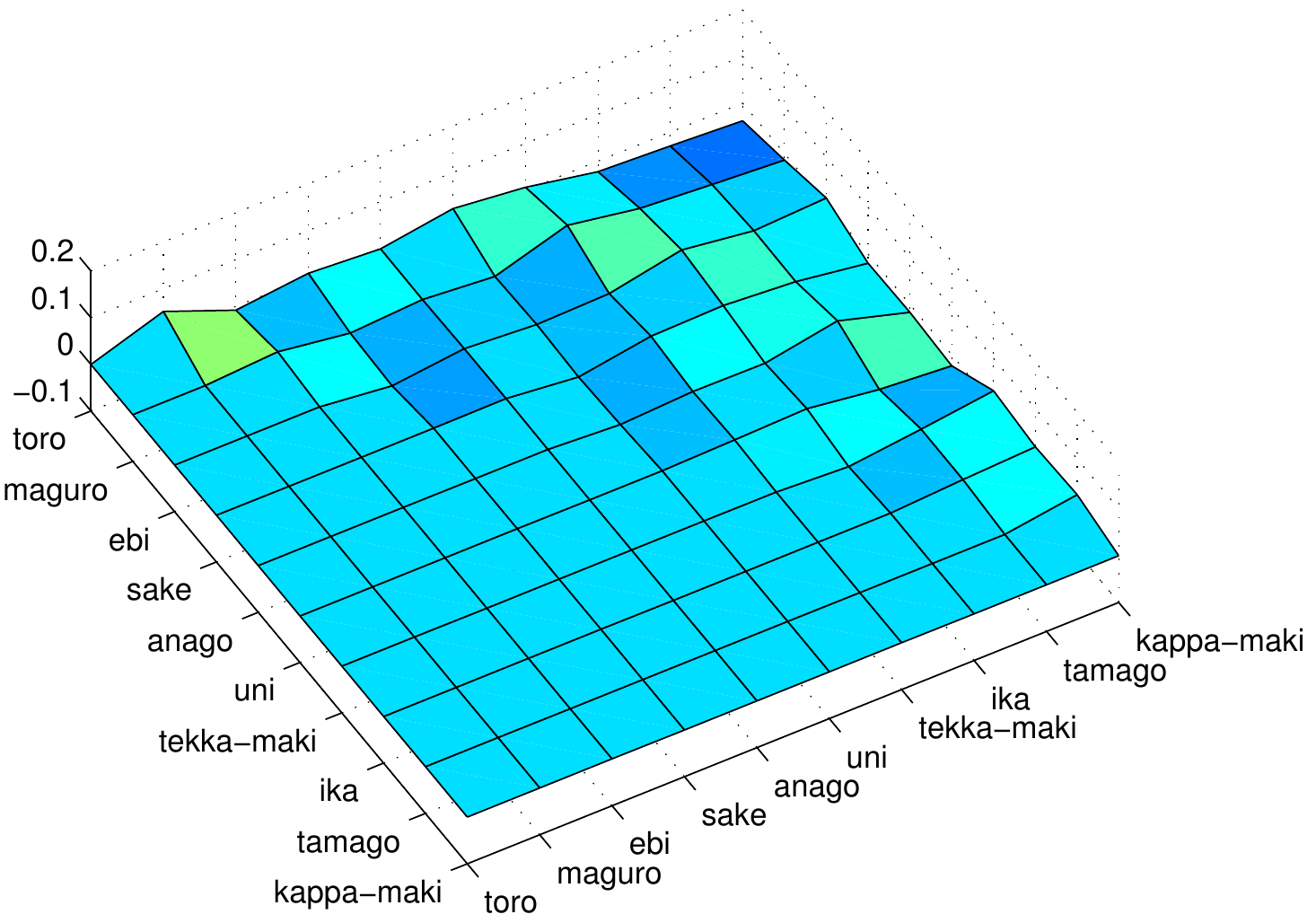}
    \label{fig:normal-deviation}
  }  
  \caption{Comparison of different models for the sushi data: each row
    shows one model with axes of plots arranged in the model's modal
    ordering. The first column shows the model's predicted pairwise
    comparison probabilities (for an item on the left axis to be
    ranked above an item on the right axis). The second column shows
    the empirical pairwise comparison probabilities . The last column
    shows the difference of the two. For clarity and because of
    symmetry, we plot one probability for each pair of items.}
  \label{fig:sushi-comparison}
  \vspace{-12pt}
\end{figure*}

\citet{Kamishima03} collected data on the rank order preferences of
restaurant customers in Japan for different types of sushi.  A
particularly interesting subset of this data are a collection of 5000
rank orders on the same 10 pieces of sushi: {\em ebi} (shrimp), {\em
  anago} (sea eel), {\em maguro} (tuna), {\em ika} (squid), {\em uni}
(sea urchin), {\em sake} (salmon roe), {\em tamago} (egg), {\em toro}
(fatty tuna), {\em tekka-maki} (tuna roll), and {\em kappa-maki}
(cucumber roll). These rankings, each provided by a unique customer,
are an example of {\it preference} data. By accurately modeling the
distribution of rankings, we can describe the distribution of
preferences over the population, and ultimately produce a succinct
depiction of collective preferences.


Azari~\etal originally showed that the Normal RUM has better model fit
on this dataset and others; we additionally demonstrate why this model
can better describe collective preferences than the Plackett-Luce and
Mallows models. We focus on the pairwise comparison
probabilities---the marginal probability that one alternative is
ranked above another alternative. Pairwise comparisons measure the
first-order accuracy of the model and have been used since the
earliest statistical ranking models~\cite{Mosteller51,Bradley52}.

Figure~\ref{fig:sushi-comparison} compares the aggregated pairwise
comparison probabilities to the empirical data, with one model in each
row, as surface plots of $m \times m$ matrices. The first column shows
the model's prediction according to maximum likelihood parameters. The
second column shows the empirical probabilities, and the third column
shows the element-wise difference. As each model implies a different
modal (most likely) ordering of the alternatives, the plots in each
row show the implied modal ordering along the axes. For the sake of
clarity, and recognizing symmetry, we show only one comparison between
each pair of items in the plots. A continuous color scale indicates
the magnitude of each value, and in the third column, a flatter plot
with more uniform color indicates a better fit between the model
predictions and empirical comparisons.


The Mallows model, shown in the top row, can only fit a surface of
probabilities derived from the single parameter $p$, which
monotonically increases for alternatives that are further separated in
its modal ranking (Figure~\ref{fig:mallows-pairwise}). 
Compared to empirically observed comparisons, the Mallows model
suffers from both systematic over- and under-estimates of pairwise
probabilities (Figure~\ref{fig:mallows-deviation}). In the second row,
the Plackett-Luce model, while more flexible than the Mallows model,
predicts comparison probabilities that are constrained by monotone
increases along its modal ordering (the axes of
Figure~\ref{fig:pl-pairwise}), and still shows significant deviations
from the data (Figure~\ref{fig:pl-deviation}).

%
%
%

The third row illustrates the predictions of the Normal RUM.  Due to
flexible variance parameters, one per alternative, the model is no
longer subject to the monotonic behavior we observed previously
(Figure~\ref{fig:normal-pairwise}), and bears a much closer
resemblance to the empirical probabilities
(Figure~\ref{fig:normal-empirical-pairwise}).  The Normal RUM is able
to fit the pairwise, empirical probabilities very closely compared to
the other models (Figure~\ref{fig:normal-deviation} is much flatter).

%
%
%
%
%

Since these models attempt to capture a symmetric distribution over
the population and not predictions for each person, the increased
explanatory power of the Normal RUM is not due to overfitting: the
model consists of only 20 parameters, while the data contains 5000
permutations of 10 alternatives.
%

\paragraph{Model Interpretation}

\begin{figure*}[t]
  \centering
  \includegraphics[trim=20mm 8mm 20mm 8mm,clip,width=0.75\textwidth]{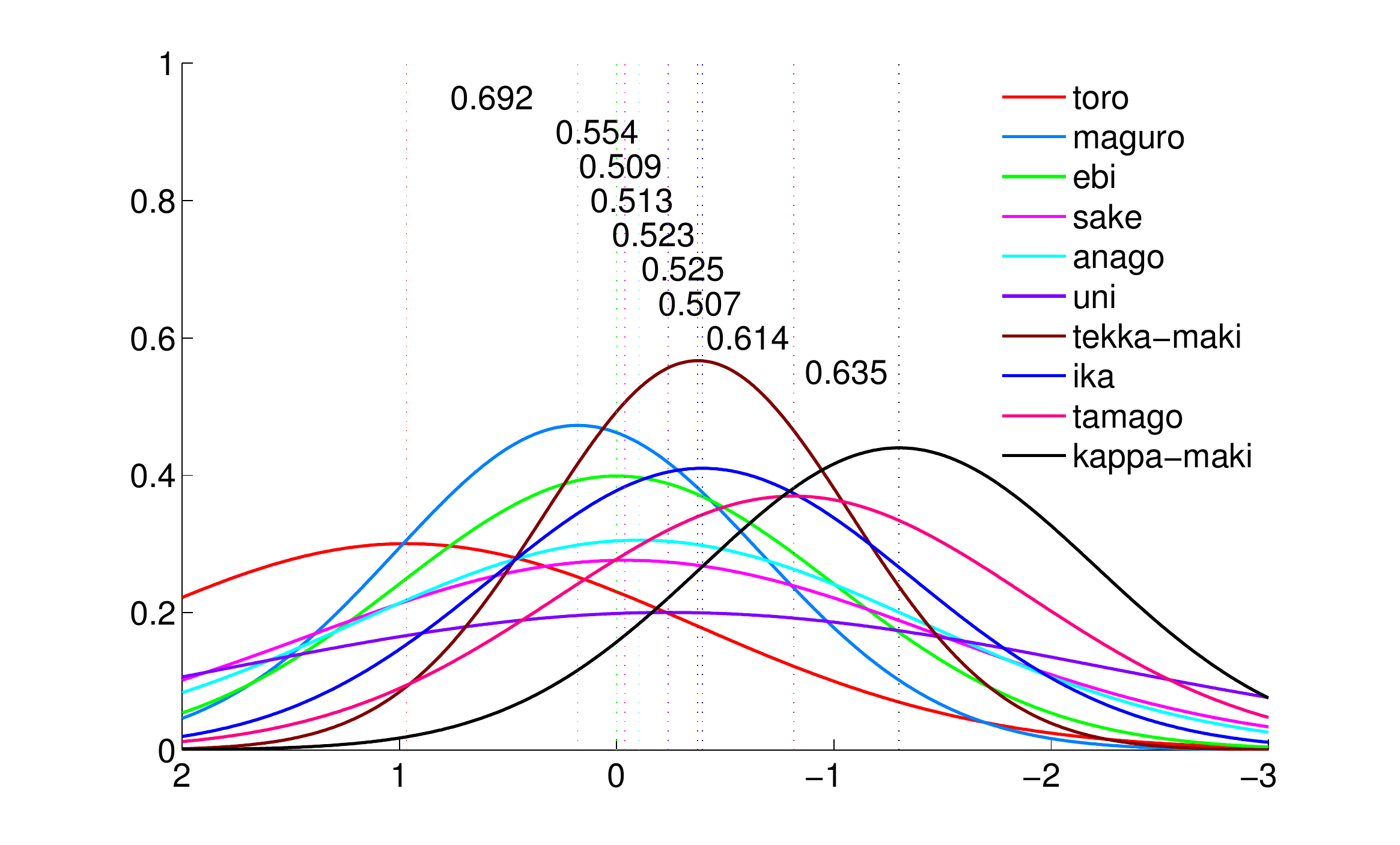}
  \caption{Distribution of random utilities for the different types of
    sushi in the estimated Normal RUM. Values show the probabilities
    of two adjacent sushi in the modal ordering to be ranked in that
    order. For example, $0.692$ is the predicted probability that {\it
      toro} is ranked ahead of {\it maguro}, and $0.554$ is the
    predicted probability that {\it maguro} is ranked ahead of {\it
      ebi}.}
  \label{fig:sushi-noise}
\end{figure*}

%

As the Normal RUM achieves a much more accurate estimate of marginal
pairwise probabilities in the data, we now turn to interpretation of
its distribution.
Figure~\ref{fig:sushi-noise} plots the estimated, random utility
distributions for each alternative. For identifiability in estimation
of the model, an arbitrary distribution ({\it maguro}) is fixed to the
standard normal distribution. Note that the x-axis is reversed, with
larger values to the left.  Under the model, each consumer's
preferences are represented by an independent draw of random values
from each of the the distributions, and ranked according to the
realized values.
%
The plot also shows the predicted, marginal pairwise probabilities
that adjacent pairs in the modal ordering are ranked according to that
ordering. 

This interpretation yields a great deal of information about the
preferences of sushi consumers. First, the most preferred ({\it toro})
and least preferred ({\it kappa-maki}) show a clear separation from
the rest of the alternatives, and are separated from their immediate
neighbors with a much greater probability than other adjacent pairs in
the ranking. This shows greater universality in the like or dislike of
these alternatives.\footnote{{\it Toro} is a fatty cut from the belly
  of the bluefin tuna that is especially highly regarded, and
  invariably commands a premium price in sushi bars. {\it Kappa-maki}
  is a perhaps unremarkable sushi of cucumber and rice.} In contrast,
preferences over adjacent items in the middle of the ranking are more
noisy and comparison probabilities are close to $0.5$ for many pairs.

The variance of the distribution for each type of sushi is also
informative.  The greatest variance is for {\it uni}, while {\it
  tekka-maki} (tuna roll) has the lowest variance, revealing a
dichotomy of preference for the former and a more consistent but
average support for the latter.\footnote{In contrast to the
  conventional {\it tekka-maki} (tuna roll), {\it uni} is a rather
  unique type of sushi made from the gonads of the sea urchin, known
  to elicit delight or disgust depending on a person's taste.}
%
%
This illustrates how the Normal RUM allows a very large set of ranking
data to be distilled into an intuitive explanation of the
population. Notably, this intepretation would not be possible with the
widely-used Mallows and Plackett-Luce models, whose parametrizations
cannot capture the differing variance across items and are thus more
inaccurate in representing the diversity of preferences. We discuss
these model deficiencies in further detail in
Appendix~\ref{sec:model-properties}.

\section{Ranking Decision Dataset}
\label{sec:ranking}


Mao \etal\shortcite{MaoAAAI13} collected data on human judgment in two
different ranking problems that were designed for the level of
difficulty to be systematically varied.
%
We briefly describe
the nature of the data below; for more details, please consult the
original paper.

\begin{figure}
  \includegraphics[width=0.45\columnwidth]{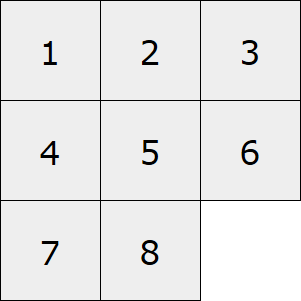}
  \hspace{0.09\columnwidth}
  \includegraphics[width=0.45\columnwidth]{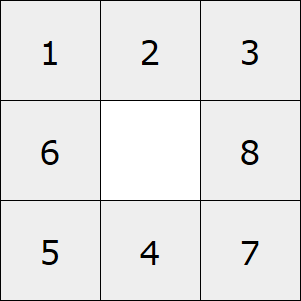}
  \caption{Two 8-puzzle states: on the left, the goal state. On the
    right, a state that requires at least 10 moves to reach the
    goal. The puzzle dataset consists of rankings of sets of four
    pictures like the one on the right.}
  \label{fig:puzzle}
\end{figure}

\paragraph{Data set 1: Sliding 8-puzzles.} 


The 8-puzzle (Figure~\ref{fig:puzzle}) consists of a square 3x3 board
with tiles numbered from 1 to 8 and an empty space. From any legal
board state, one solves the puzzle by sliding one tile after the next
into the empty space, seeking to obtain a board state where the
numbers are correctly ordered from top to bottom and left to right.
Each movement of a single tile counts as one ``move'', and one goal is
to solve the puzzle in as few moves as possible.

In the 8-puzzle dataset, users ranked four instances of 8-puzzle game
states by the least number of moves from the solution state, from
closest to farthest away. Choosing a sequence of numbers, such as $(7,
10, 13, 16)$, and generating a set of four random puzzles solvable in
a corresponding number of moves as computed by $A^*$ search produced
ranking data at a specific, consistent difficulty.
By fixing the difference between the numbers but varying the overall
distance to the goal, the puzzles become harder or easier to rank
relative to each other.


\begin{figure}
  \includegraphics[width=0.45\columnwidth]{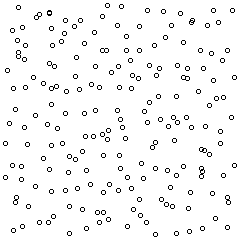}
  \hspace{0.09\columnwidth}
  \includegraphics[width=0.45\columnwidth]{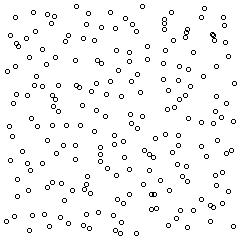}
  \caption{Two pictures of dots: on the left, a picture with 200
    dots. On the right, a picture with 208 dots. The dots dataset
    consists of rankings over sets of four pictures such as these.}
  \label{fig:dots}
\end{figure}

\paragraph{Date set 2: Dot fields.} The problem of counting
pseudo-randomly distributed dots in images has been suggested as a
benchmark task for human
computation~\cite{HortonDots}. \citet{Pfeiffer2012} used the task of
comparing these dot fields as a proxy for noisy comparisons of items
in ranking tasks.
In the dots dataset, people ranked four pictures with many more dots
than could be manually counted.  For example, Figure~\ref{fig:dots}
shows pictures with 200 and 208 dots, respectively.  Varying the
difference in the number of dots between pictures allows adjustment of
the difficulty of the task.

\vspace{5pt} In both domains, there are four levels of difficulty,
each with 40 separate sets of four alternatives to rank. Around 20
ranking orders are elicited from people for each set, providing a
total of 3,200 rankings for each of the two ranking problems.  In
contrast to the sushi data described in Section~\ref{sec:sushi}, these
ranking datasets do not capture different preferences across
individuals. Instead, every user has the same preferred ranking over
the data, but variance arises from {\em imperfect or noisy
  perception}. Thus, by learning a distribution of rankings over the
data, we can describe decision-making ability and cognitively
difficult comparisons across the collective population.


\paragraph{Model Interpretation}
\label{sec:ranking-noise}

\begin{figure}[t!]
  \centering 
  \subfloat[Adjacent pairwise probabilities.] {
    \includegraphics[width=\figurewidth]{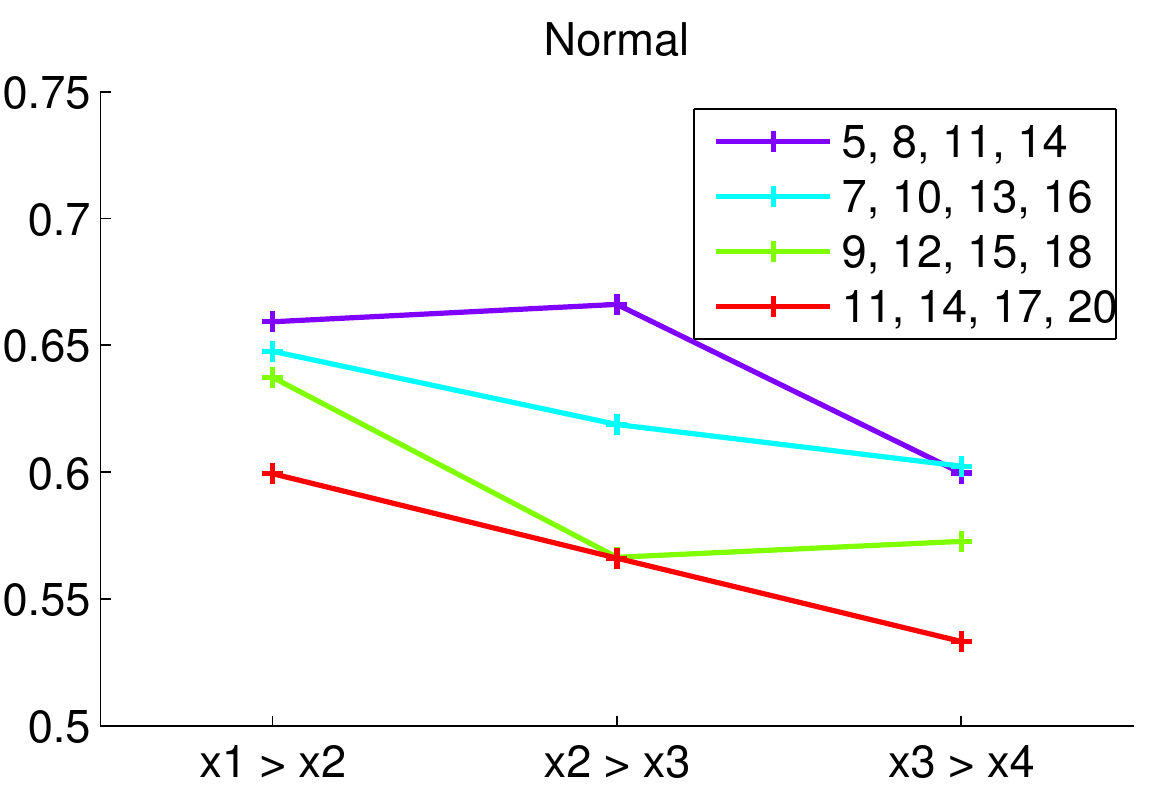}  
    \label{fig:puzzle-probs}
  }
  \\
  \vspace{-10pt}
  \subfloat[5, 8, 11, 14 moves: top line of {\bf (a)}.] {
    \includegraphics[width=\figurewidth]{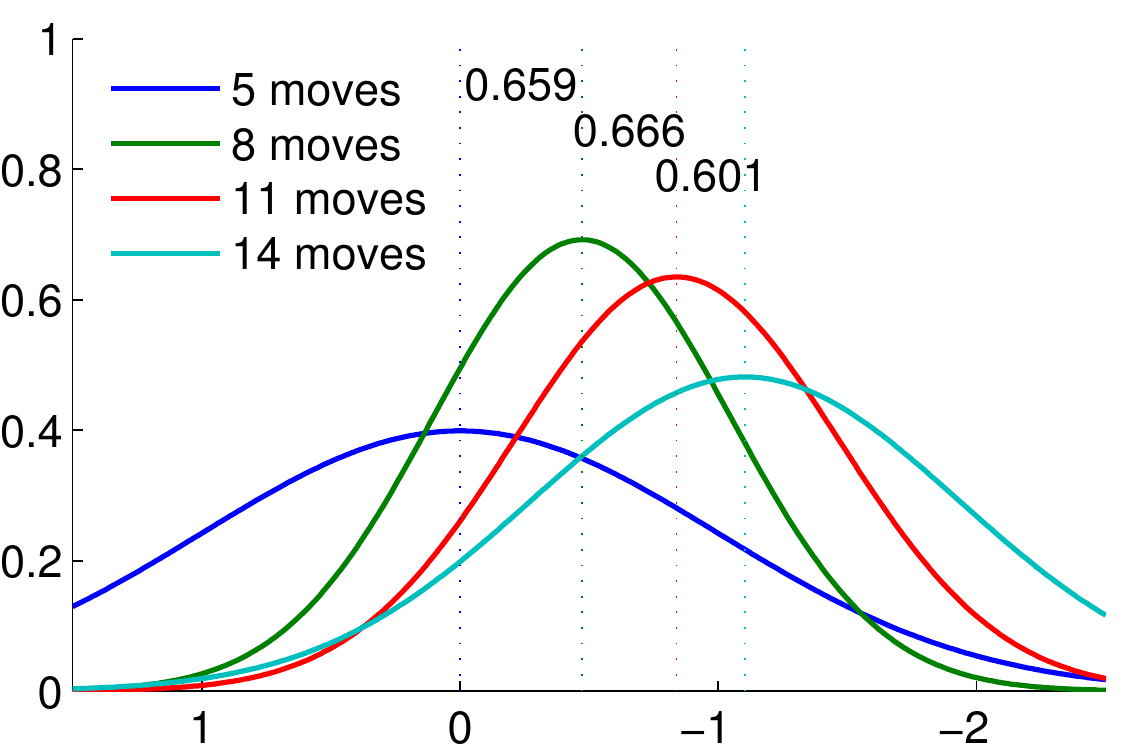}
    \label{fig:puzzle-5}
  }
  \\
  \vspace{-10pt}
  \subfloat[11, 14, 17, 20 moves: bottom line of {\bf (a)}.]  {
    \includegraphics[width=\figurewidth]{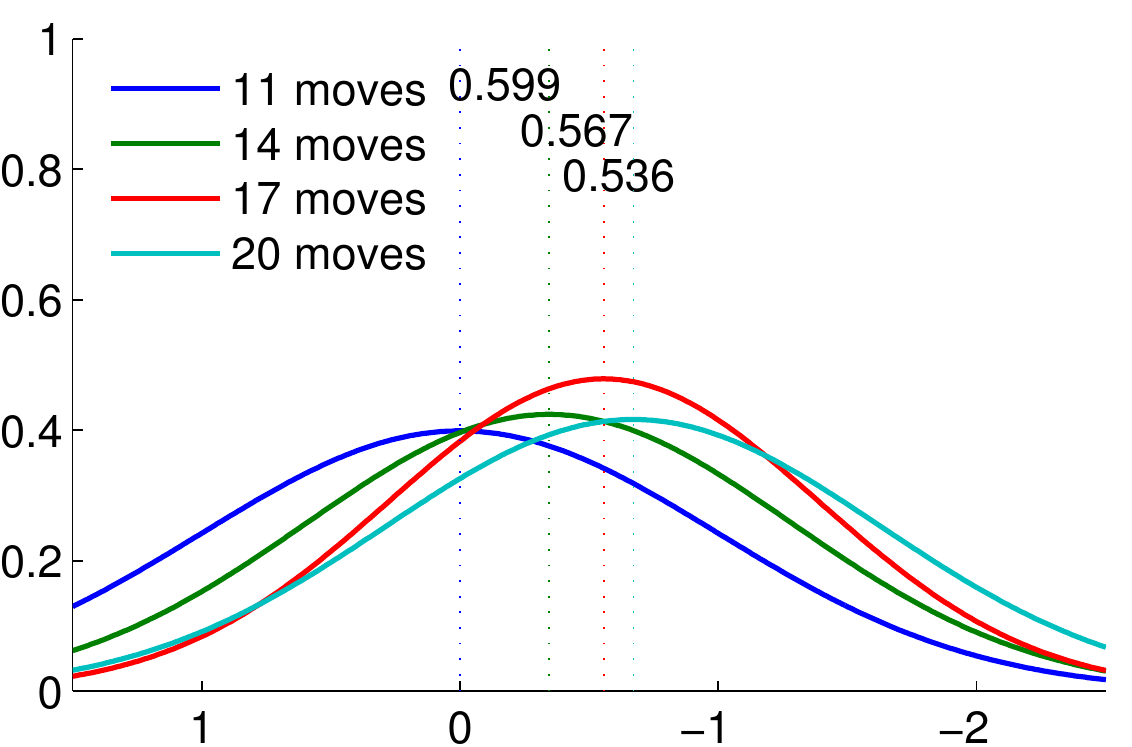}
    \label{fig:puzzle-11}
  }
  \caption{Aggregate fitted results for the normal model on the
    8-puzzle rankings. Dashed lines indicate mean strength. Three
    numbers in each plot show probabilities of adjacent pairwise
    probability implied by the model.}
  \label{fig:puzzle-noise}
\end{figure}

\begin{figure}[t!]
  \centering 
  \subfloat[Adjacent pairwise probabilities.] {
    \includegraphics[width=\figurewidth]{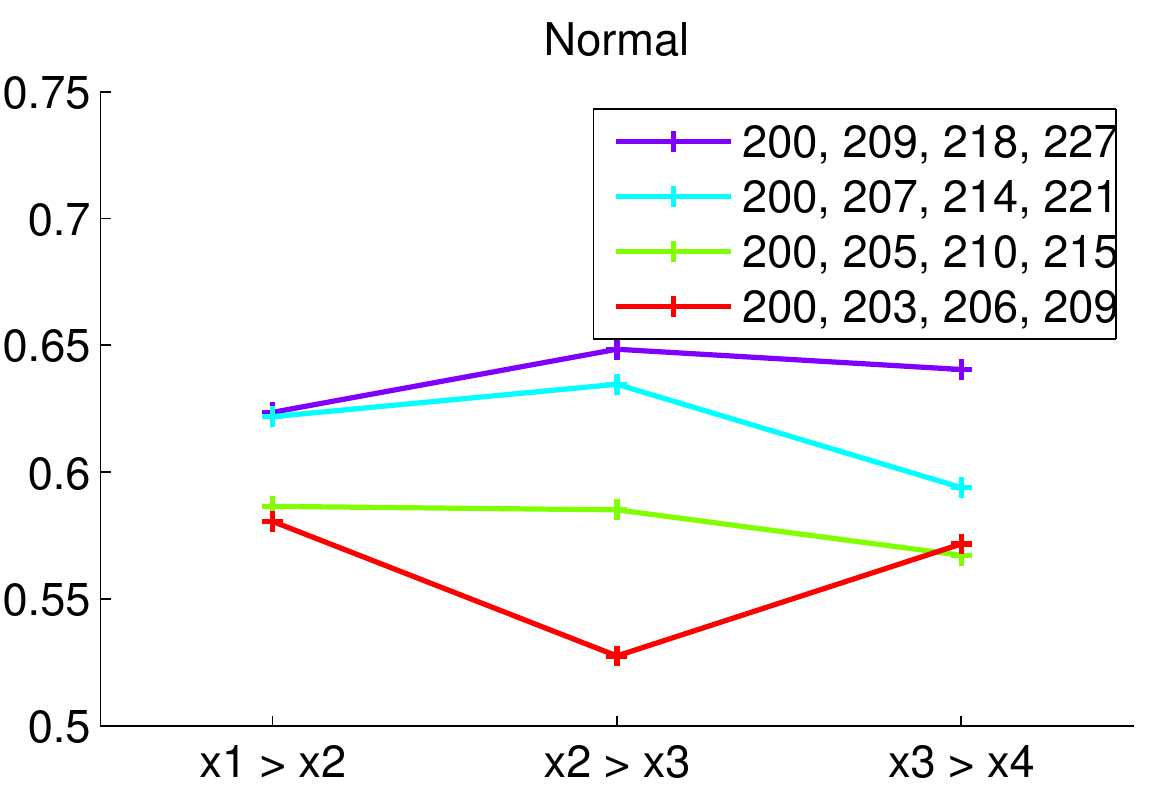}
    \label{fig:dots-probs}
  }
  \\
  \vspace{-10pt}
  \subfloat[200, 209, 218, 227 dots: top line of {\bf (a)}.] {
    \includegraphics[width=\figurewidth]{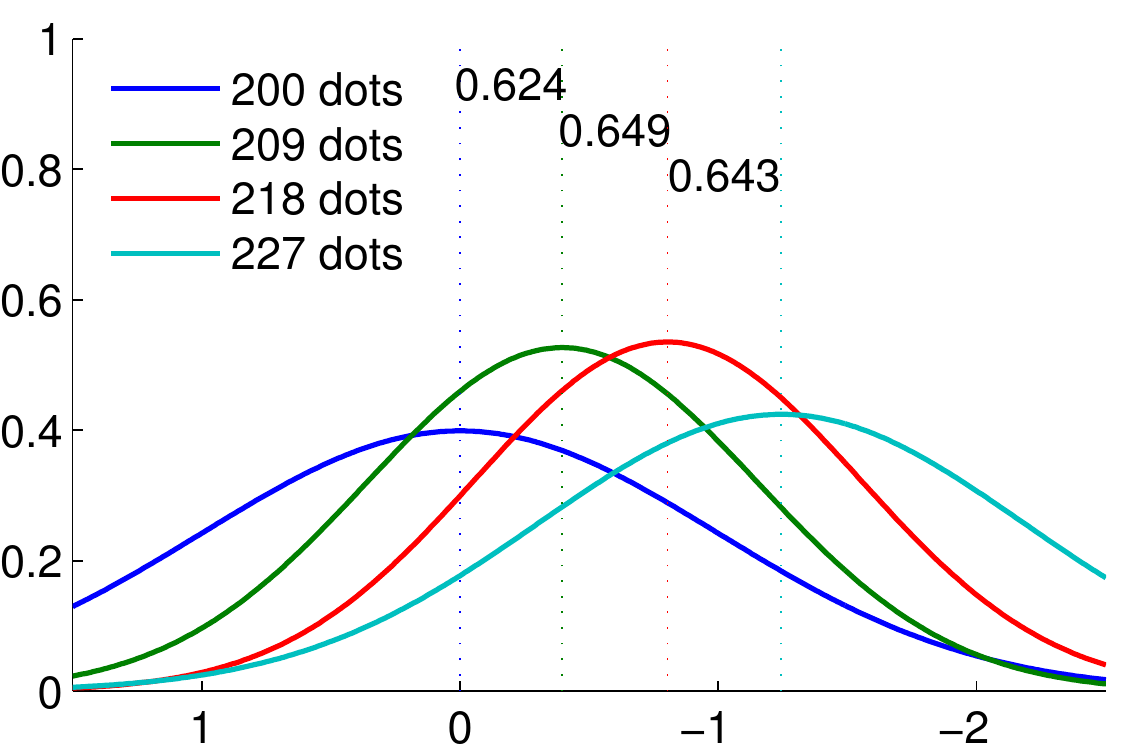}
    \label{fig:dots-9}
  }
  \\
  \vspace{-10pt}
  \subfloat[200, 203, 206, 209 dots: bottom line of {\bf (a)}.]  {
    \includegraphics[width=\figurewidth]{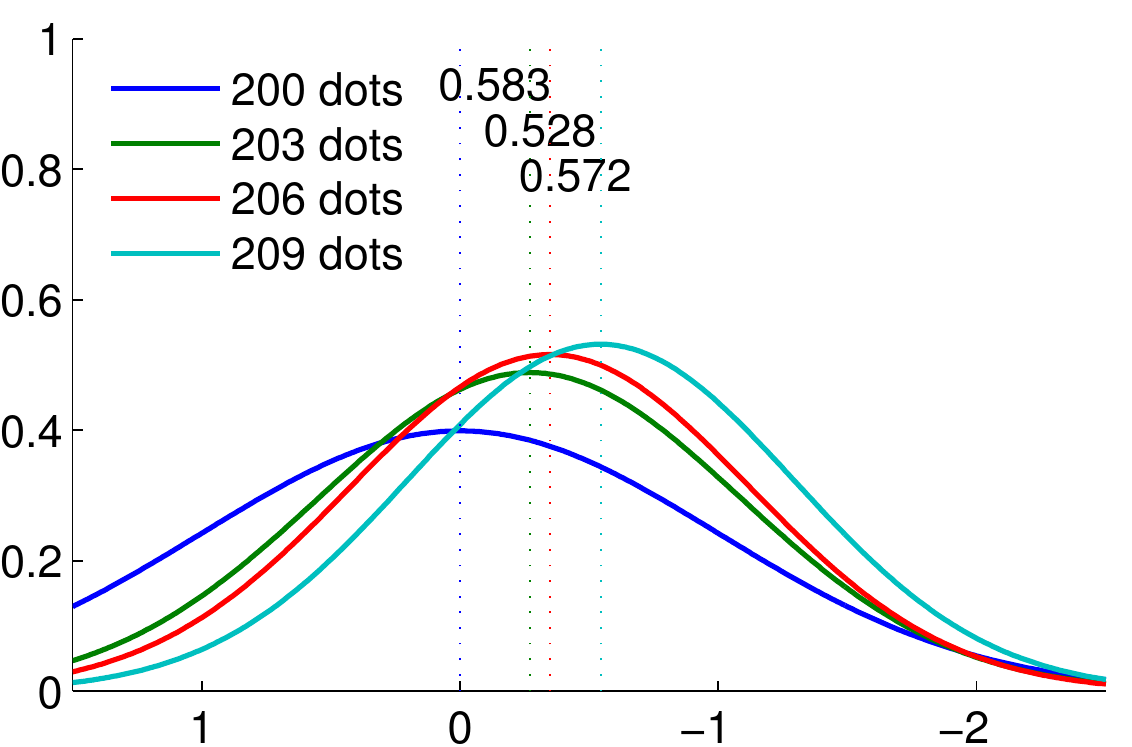}
    \label{fig:dots-3}
  }
  \caption{Aggregate fitted results for the normal model on the dots
    image rankings. Dashed lines indicate mean strength. Three numbers
    in each plot show probabilities of adjacent pairwise probability
    implied by the model.}
  \label{fig:dots-noise}
\end{figure}


As in the sushi dataset, the Normal RUM provides the best
approximation of empirical comparison probabilities on this data. We
focus on the estimated parameters to gain an understanding of human
perceptive judgment in these two domains.

Figures~\ref{fig:puzzle-noise} and~\ref{fig:dots-noise} show the
properties of the estimated random utility distributions for each
problem. The first graph shows the pairwise comparison probabilities
for adjacent alternatives at each level of difficulty. The second and
third plots show the estimated random utility distributions for the
easiest and hardest ranking problems, respectively.
%
In each case, the distributions are scaled so that the mean and
variance of the highest-ranked item (closest puzzle to the goal or
lowest number of dots) has the standard normal distribution.
%

In both problems, adjacent pairs of items become harder to rank as
difficulty increases, with comparison probability decreasing toward to
$0.5$---shown by the progression of successively lower line segments
in Figures~\ref{fig:puzzle-probs} and \ref{fig:dots-probs}. Yet, it is
particularly interesting how the varying difficulty of the two
different problems affects users' judgments, as revealed from the
estimated model.
%
%
%
%
For the 8-puzzle, the probability of correctly ordering adjacent
puzzles slopes downward from left to right, showing that is harder to
compare two puzzles that are further from the goal state than two that
are closer. This is naturally explained by observing that it is
easier, for example, to tell which of two puzzles that are 7 and 10
moves from the solution is closer, than two puzzles that are 13 and 16
moves from the solution. All levels of difficulty in the problem
exhibit this property.

For the fields of dots, there is a different behavior as difficulty
increases. At the easier levels of difficulty, the probability of
ranking adjacent alternatives in the correct order does not
significantly slope from left to right for the three adjacent
pairs. This is quite different from the 8-puzzle setting, but is
naturally explained considering that dot fields do not become harder
to rank when the difference of dots is a similar percentage of the
total (around 200 for all pictures). However, in the most difficult
setting, with a difference of only 3 dots between pictures, there is a
marked drop in accuracy for the intermediate two pairs of
pictures.

In the distribution over rankings, we see a similar pattern to the
sushi dataset: namely, there is more certainty about the pictures with
the least and most number of dots (resp. stronger preference of the
favorite and least favorite sushi) than among the intermediate
choices. In the dots data, the lower certainty arises from increased
perceptive error between the extremes, while in the sushi data this
manifests from more varied preferences apart from the favorite and
least favorite. The Normal RUM captures both of these cases, allowing
it to describe both rankings of preference and rankings of imperfect
perception.

As each set of parameters is generated from 800 rankings and the
patterns we observe are consistent across the difficulty levels for
both problems, we believe these results are robust. Compared with the
parameters estimated for sushi data, the variance of the noise
distributions for alternatives is more uniform.
However, the expressiveness of the Normal RUM allows similar variances
across alternatives to be interpreted as more or less uniform
perceptive error in ranking, rather than a necessary limitation of the
model (as is the case with Plackett-Luce: see
Appendix~\ref{sec:model-properties}).



\section{Discussion}
\label{sec:discussion}

Our results illustrate the importance of flexible and expressive
models for fitting human ranking data, encompassing both variation
across {\it population preference} as well as errors arising from {\it
  imperfect perception}. Commonly used but restrictive models such as
Mallows and Plackett-Luce are insufficient to capture the various
patterns encountered in human perception. However, an appropriately
descriptive model such as the Normal RUM allows for intuitive
interpretation of both types of data in a more nuanced way.

%
Such models allow us to gain new understanding and intuition about
real data. We are able to see clear preferences for certain
alternatives by users, as well as which choices present more
contention or uncertainty and which are uncontroversial. We can also
see how users' perception and decision making is affected by harder
and easier tasks, and for which alternatives they can make judgments
that are either more certain or more ambiguous.
Our approach also highlights the importance of detailed evaluation
techniques that reveal the quality of a model's fit to data in a
natural way, such as looking at pairwise comparison probabilities. Our
results clearly show how classical models fall short in representing
human ranking data.

%



Our results motivate several areas of future work. The Normal RUM and
other flexible random utility models are promising for discovering
interesting patterns in preferences across a collective group of
people. In our case, estimating such a model distilled several
thousand rankings into a much more concise representation. We believe
that this type of analysis facilitates a more natural understanding of
collective preferences over simple rank aggregation.  At the same
time, we foresee that a better understanding of the difficulty and
variance in human judgment problems
can motivate the design of better user interfaces and crowdsourcing
systems. Using any ranking data, one can explore errors and variance
when human users are asked to make comparisons that are noisy or
uncertain.

Our hope is that the analysis and methods in this work will motivate
more insightful descriptions of collective human preferences as well
as a better understanding of decision making in the design of systems
involving human agents.


%
%




\small
\bibliographystyle{aaai}
\bibliography{abb,combined,ultimate}

\newpage
\appendix

\section{Model Properties}
\label{sec:model-properties}

Sections~\ref{sec:sushi} and~\ref{sec:ranking} showed that the Normal
RUM allows a better interpretation of several data sets than the
classical Mallows and Plackett-Luce models, because of its advantages
in correctly capturing comparison probabilities over all pairs of
alternatives. Here, we show analytically why this generally can be
true for any data set, due to inherent restrictions in the classical
models.

By estimating a ranking model $\mathcal{M}$ (i.e., by using maximum
likelihood), we obtain parameters $\hat{\theta}$ implying a
distribution on rank orders as well as a marginal distribution over
pairwise comparisons. We generally expect that the probability of
users ranking one particular item above another will be only
marginally affected by the presence of other choices,
so we can evaluate the suitability of a model by how well it
approximates all marginal pairwise probabilities. More generally,
pairwise comparison data has a long history of use in learning
rankings~\cite{Mosteller51,Bradley52} and is generally viewed as more
robust than cardinal data~\cite{Ammar11}.

Under the Normal RUM, the probability that a particular alternative
$a_j$ is preferred to another item $a_k$ is
\begin{equation}
  \Pr_{\text{Normal}}(a_j \succ a_k \mid \boldsymbol{\mu}, \boldsymbol{\sigma} ) = 
  \Phi\left( ({\mu_j - \mu_k})/\sqrt{\sigma_j^2 +\sigma_k^2} \right)
  \label{eq:normal-prob}
\end{equation}

where $\Phi$ is the CDF of the standard normal distribution.
Figure~\ref{fig:sushi-comparison}, showed that this two-parameter
model closely approximated all pairwise comparisons in a large set of
empirical preference data.


\paragraph{Limitations of the Mallows Model}



Under the Mallows model, the noise parameter $p$ is also the
probability that adjacent pairs in the reference ranking $\sigma^*$
are ranked in the same order. More generally, any two items separated
by a fixed distance in $\sigma^*$ are ranked correctly with the same
probability $\Pr(a_{\sigma^*(k)} \succ a_{\sigma^*(l)})$ for positive
$z = l - k$, which can be shown to be
\begin{equation}
  \Pr_{\text{Mallows}}(a_{\sigma^*(k)} \succ a_{\sigma^*(l)} \mid \phi, \sigma^*) = \frac{\sum_{z=1}^c z \phi^{z-1}}
  {\left(\sum_{z=1}^c \phi^{z-1}\right)
    \left(\sum_{z=0}^c \phi^{z}\right)}
\label{eq:mallows-pairwise}
\end{equation}
with $\phi = (1-p)/p$.
As shown in Figure~\ref{fig:mallows-pairwise}, this is monotone
increasing in $c$ for a given $\phi$. In the context of rankings from
human input, this rather strong assumption is easily violated, such as
when agents have more certain comparisons over adjacent pairs in one
part of the ranking than the other. In the case of the sushi data and
the most difficult dots ranking problem, this occurred at endpoints of
the ranking with more uncertainty in the middle. Hence, while popular
and extended in many ways as outlined in
Section~\ref{sec:model-extensions}, the Mallows model is inherently
rather restrictive.

\paragraph{Limitations of The Plackett-Luce Model}

Under the Plackett-Luce model, a rather severe restriction
is the fixed variance of the Gumbel distribution for the random
utility across alternatives. 
The marginal probability
that alternative $a_j$ ranked is higher than alternative $a_k$ is
\begin{equation}
  \Pr_{\text{PL}}(a_j \succ a_k \mid \boldsymbol{\mu}) = 1/\left(1 + e^{-(\mu_j - \mu_k)}\right)
  \label{eq:logit-pairwise}
\end{equation}

%
%
Since the logistic sigmoid $g(x) = 1/(1 + e^{-x})$
is strictly monotonically increasing, there is a strict limitation in
which pairwise probabilities for ordering items in a particular way
are determined only by the difference in their strength values.
Specifically, for any strictly monotone increasing function $f(x)$, we can show that for any fixed $\mu_j$,
\begin{equation}
  \mu_k > \mu_k' \Longleftrightarrow f(\mu_j - \mu_k) < f(\mu_j - \mu_k')
  \label{eq:monotone}
\end{equation}
This behavior is exemplified in Figure~\ref{fig:pl-pairwise}, where
the comparison probabilities monotonically increase along the axes,
and emerges in all fixed-variance (one-parameter) random utility
models.

As we have seen, this assumption is rather strong, and any RUM with
this property cannot capture the notion that it may be less certain to
compare some particular alternative (such as {\em uni}) versus others,
and hence it would be sensible for the variance of utility for that
alternative to change rather than the mean value.




\end{document}